\documentclass{ws-ijmpb}
\usepackage{xcolor}

\begin{document}

\markboth{Lara Ul{\v c}akar and Anton Ram{\v s}ak}
{Effects of noise on fidelity in spin-orbit qubit transformations}

%
\catchline{}{}{}{}{}
%

\title{Effects of noise on fidelity in spin-orbit qubit transformations }

\author{Lara Ul{\v c}akar}

\address{J. Stefan Institute, Ljubljana, Slovenia\\
lara.ulcakar@ijs.si}

\author{Anton Ram\v{s}ak}

\address{Faculty of Mathematics and Physics, University of Ljubljana, Ljubljana, Slovenia\\
J. Stefan Institute, Ljubljana, Slovenia\\
anton.ramsak@fmf.uni-lj.si}

\maketitle

\begin{history}
\end{history}

\begin{abstract}
We analyse non-adiabatic non-Abelian holonomic transformations of spin-qubits confined to a linear time dependent harmonic trap with time dependent Rashba interaction. For this system exact results can be derived for spin rotation angle which also enables exact treatment of white gate-noise effects. We concentrate in particular on the reliability of  cyclic transformations quantified by fidelity defined by the probability that the qubit after one full cycle remains in the ground-state energy manifold.
The formalism allows exact analysis of  spin transformations that optimise final fidelity. Various examples of time dependent fidelity probability distributions are presented and discussed.
\end{abstract}


\section{Introduction}
Spintronics, as a new branch of electronics, is a quantum information technology promising  better performance with smaller power consumption.\cite{Wolf2001,Zutic2004,Rashba2007} The spin of electrons plays the central role\cite{Awschalom2013} and the main challenge is to manipulate the spin of a single electron precisely and locally. Employing magnetic fields, a natural way of spin rotation, usually cannot be applied locally in a small region so other mechanisms should be applied. A possible such solution is to use semiconductor heterostructures\cite{Winkler2003,Engel2007} with spin-orbit interaction (SOI) and particularly strong Rashba interaction\cite{Dresselhaus1955,Rashba1960} that can be tuned externally using voltage gates.\cite{Nitta1997,Schapers1998,Nitta1999,Schliemann2003,Wunderlich2010,gomez12,pavlowski16,pavlowski16b,fan16} 

Recently a simple scheme for the spin-qubit manipulation was proposed in which an electron is driven along a linear quantum wire with time dependent spin-orbit interaction, tuned by external time-dependent potential.\cite{Cadez2013,Cadez2014} One limitation of such linear systems is posed by fixed axis of spin rotation, but it can  be eliminated in quantum ring structures, exhibiting a rich range of phenomena.\cite{Buttiker1983,Fuhrer2001,Aronov1993,Qian1994,Malshukov1999,Richter2012,Nagasawa2013,Saarikoski2015} For quantum ring structures consisting of a narrow ring with superimposed time dependent harmonic trap and controllable time dependent Rashba interaction exact solutions were presented most recently.\cite{kregar16,kregar16b} 

In linear as well as in ring systems controlled by external gates there are several possible sources of noise which can not be avoided. In particular, noise can be induced
due to fluctuating electric fields, caused by the piezoelectric phonons\cite{sanjose06,sanjose08,huang13,echeveria13} or due to phonon-mediated instabilities in molecular systems with phonon assisted potential barriers, which introduce noise in the confining potentials.\cite{mravlje06,mravlje08} For qubits realised as spin of electrons  carried by surface acoustic waves the noise can be caused by the electron-electron interaction.\cite{rejec00,jefferson06,giavaras06} 
Since exact solutions for  qubit manipulation scheme considered here are possible, the analysis of  environment effects can for some  sources of noise be performed analytically.\cite{ulcakar17}

The paper is organised as follows. After the introduction is in Section~2  presented the model  where also a brief overview of the exact solution together with the analysis of effects due to white noise  is revealed.  Section~3 is devoted to the fidelity of qubit transformations. The derivation of influences of noise on fidelity is presented in detail and explicit examples are given. Results are summarised in Section~4.

\section{Model, exact solution and white noise}

We consider an electron in a quantum wire confined in a harmonic trap.\cite{Cadez2013,Cadez2014} The centre of such one-dimensional quantum dot,  $\xi(t)$, can be arbitrarily translated along the wire by means of time dependent external electric fields. Spin-orbit Rashba interaction couples the electron spin with orbital motion, resulting in the Hamiltonian
\begin{equation}\label{H}
H(t)=\frac{p^{2}}{2m^{*}}I+\frac{m^{*}\omega^{2}}{2}[x-\xi(t)]^{2}I+\alpha(t)p\, \mathbf{n}\boldsymbol{\cdot\sigma},
\end{equation}
where $m^{*}$ is the electron effective mass, $\omega$ is the frequency
of the harmonic trap, $\alpha(t)$ is the strength of spin-orbit interaction, possibly time 
dependent due to appropriate time dependent external electric fields.  The spin rotation axis $\mathbf{n}$ is fixed and depends 
on the crystal structure of the quasi-one-dimensional material used and 
the direction of the applied electric field.\cite{nadjperge12} $\boldsymbol{\sigma}$
and $I$ are Pauli spin matrices and unity operator in spin space, respectively, and $p$ is the momentum operator. Exact solution of
the time dependent Schr{\"o}dinger equation corresponding to the Hamiltonian equation~\eqref{H} is given 
by\cite{Cadez2014}
\begin{eqnarray}\label{psi}
|\Psi_{ms}(t)\rangle&=&e^{-i[\theta(t) I+\phi(t) \mathbf{n}\cdot\boldsymbol{\sigma}/2)]}
\mathcal{A}_{\alpha}\mathcal{X}_{\xi}|\psi_{m}(x)\rangle|\chi_{s}\rangle,
\\
\theta(t)&=&{\omega_{m}t+\phi_{\alpha}(t)+{\phi_{\xi}(t)}+m^{*}\dot{a}_{c}(t)a_{c}(t)/\omega^{2}},\\
\mathcal{A}_{\alpha}&=&
e^{-i\dot{a}_{c}(t)p\mathbf{n}\cdot\boldsymbol{\sigma}/\omega^2}
e^{-im^{*}a_{c}(t)x\mathbf{n}\cdot\boldsymbol{\sigma}},\\
\mathcal{X}_{\xi}&=&e^{im^{*}[x-x_{c}(t)]\dot{x}_{c}(t)}e^{-ix_{c}(t)p}I.
\end{eqnarray}
Here $\psi_{m}(x)$ represents the $m$-th eigenstate of a harmonic oscillator with eigenenergy $\omega_{m}=(m+1/2)\omega$ and $|\chi_{s}\rangle$ is spinor of the electron in the eigenbasis of operator $\sigma_z$.
The phase $\phi_{\xi}(t)=-\int_{0}^{t}L_{\xi}(t')\mathrm{d}t'$ is the coordinate action
integral, where $L_{\xi}(t)=m^{*}\dot{x}_{c}^{2}(t)/2-m^{*}\omega^{2}[x_{c}(t)-\xi(t)]^{2}/2$
is the Lagrange function of a driven harmonic oscillator and $x_{c}(t)$ is the
solution to the equation of motion of a classical driven oscillator
\begin{equation}\label{eq:xoscillator}
\ddot{x}_c(t)+\omega^{2}x_{c}(t)=\omega^{2}\xi(t).
\end{equation}
Another phase factor is the SOI action integral phase 
$\phi_{\alpha}(t)=-\int_{0}^{t}L_{\alpha}(t')\mathrm{d}t'$,
with $L_{\alpha}(t)=m^{*}\dot{a}_{c}^{2}(t)/(2\omega^2)-m^{*}[a_{c}(t)-\alpha(t)]^{2}/2+m^{*}\alpha^2(t)/2$
being the Lagrange function of another driven oscillator, satisfying
$\ddot{a}_{c}(t)+\omega^{2}a_{c}(t)=\omega^{2}\alpha(t).
$

In this paper we consider  particularly interesting cyclic transformations with periodic drivings $\xi(T)=\xi(0)$ and $\alpha(T)=\alpha(0)$ with zero values and time derivatives of responses $x_c$ and $a_c$  at times $t=0$ and $t=T$.  The spin-qubit is for such drivings  rotated around $\mathbf{n}$ by the angle $\phi=-2m^{*}\int_{0}^{T}\dot{a}_{c}(t')\xi(t')\mathrm{d}t'$.\cite{Cadez2014}

We assume noise in the driving function $\xi(t)=\xi^{0}(t)+\delta\xi(t)$ consisting of ideal driving part without noise
$\xi^{0}(t)$ with superimposed  stochastic part with vanishing mean  $\langle\delta\xi(t)\rangle=0$. We consider
the Ornstein-Uhlenbeck coloured noise\cite{wang45,masoliver92} characterized by the autocorrelation function
$\langle\delta\xi(t')\delta\xi(t'')\rangle={{\sigma_{\xi}^2} \over {2\tau_\xi} }e^{ |t'-t''|/\tau_\xi}$,  with noise intensity $\sigma^2_\xi$ and correlation time $\tau_\xi$. 
A general solution of equation~\eqref{eq:xoscillator} $x_{c}(t)$ with 
$x_c(0)=\xi^0(0)$ and ${\dot{x}_c}(0)=0$ is
given by 
\begin{equation}\label{eq:harmoicsolution}
x_{c}(t)=\xi^0(0)+\omega\int_{0}^{t}\sin[\omega(t-t')]\xi(t'){\rm d}t',
\end{equation}
which due to the noise term $\delta \xi$ is normally distributed
with the variance evaluated as equal-times autocorrelation function,
\begin{equation}
\sigma_x^2(t)=\omega^2\lim_{\Delta t \to0} \langle\int_{0}^{t}\sin[\omega(t-t')]\delta\xi(t'){\rm d}t' \int_{0}^{t+\Delta t}\sin[\omega(t-t'')]\delta\xi(t''){\rm d}t'' \rangle.
\end{equation}
For the Ornstein-Uhlenbeck noise considered here the integrals can be evaluated exactly. Nevertheless, here we
consider only the white noise limit where $\tau_\xi \to 0$ and $\langle\delta\xi(t')\delta\xi(t'')\rangle={\sigma_{\xi}^2}\delta(t'-t'')$ leading to the variances
\begin{equation}\label{1d}
\sigma_x^2(t)=\frac{1}{4}\omega\sigma_{\xi}^{2}\left(2\omega t-{{\sin{2\omega t}}}\right)
\quad {\rm and}\quad
\sigma_{\dot{x}}^{2}(t)=\frac{1}{4}\omega^{3}\sigma_{\xi}^{2}\left[2\omega t+\sin(2\omega t)\right],
\end{equation}
corresponding to  ${x}_{c}(t)$ and $\dot{x}_{c}(t)$, respectively.

Additionally to the coordinate noise is also normally distributed noise in  SOI
driving function  $\alpha(t)=\alpha^{0}(t)+\delta\alpha(t)$, where $\alpha^{0}(t)$ is ideal noiseless driving. SOI noise $\delta\alpha(t)$ is similar to the previous case of spatial driving and is again of the Ornstein-Uhlenbeck type of autocorrelation function $\langle\delta\alpha(t')\delta\alpha(t'')\rangle$ with noise intensity $\sigma^2_\alpha$ and correlation time in the white noise limit $\tau_\alpha \to 0$,
leading to the time-dependent  variances 
$\sigma_{a}^{2}(t)=           {  (\sigma_\alpha/  \sigma_{\xi})^{2}} \sigma_{x}^{2}(t)$ and
$\sigma_{\dot{a}}^{2}(t)=         { ( \sigma_\alpha/  \sigma_{\xi})^{2}} \sigma_{\dot{x}}^{2}(t)$
for $a_{c}(t)$ and $\dot{a}_{c}(t)$,  respectively.

\section{Fidelity of noisy qubit transformations}

As an example of effects of noise to spin-qubit transformations we consider driving corresponding to the class of circular paths in two dimensional coordinate-SOI space ${\cal{C}}_{\rm ad} \sim \alpha^0[\xi^0]$, 
\begin{equation}
\xi^0(t) = \xi_0 \cos( \omega t/n) \quad {\rm and} \quad \alpha^0(t) = \alpha_0 \sin( \omega t/n), \label{krog}
\end{equation}
where $n\ge2$ is integer, and the period of the transformation is $T=2\pi n/\omega$.  Periodic responses  represent contours ${\cal{C}}\sim a_c^0[\xi^0]$, where 
\begin{equation}
a^0_c(t)=\alpha_0\frac{n \left[n \sin \left( \omega t/n\right)-\sin ( \omega t )\right]}{n^2-1},
\end{equation}
with the phases  given by the area in the coordinate-SOI plane, 
\begin{eqnarray}
\phi_{\rm ad}&=&-2m^{*}\int_{0}^{T}\dot{\alpha}^0(t')\xi(t')\mathrm{d}t'=-2m^*  \!\oint_{{\cal C}_{\rm ad}} \!\!\!\!\! \alpha^0[\xi^0]  {\rm d}\xi=-2\pi m^*\xi_0\alpha_0,\label{phiAd}\\
\phi^0 &=&-2 m^*  \!\oint_{{\cal C}}  a_c^0[\xi^0]  {\rm d}\xi={ n^2\over n^2 - 1}  \phi_{\rm ad}.\label{fi}
\end{eqnarray}
The adiabatic angle $\phi_\mathrm{ad}$ corresponds to the one when circular driving is of type $n\to\infty$. Transformation angle $\phi=\phi^0+\delta \phi$ is due to the noise distributed normally around the mean $\phi^0$, with the variance after one cycle given by\cite{ulcakar17}
\begin{equation}\label{sigma0}
\frac{\sigma_{\phi,n}^{2}}{\phi_{\rm ad}^2}= \frac{n(1+n^{2})}{\pi(n^{2}-1)^{2}}\frac{\omega\sigma_{\xi}^2}{ \xi_0^2}  +
\frac{2 n^{3}}{\pi(n^{2}-1)^{2}}\frac{\omega\sigma_{\alpha}^2}{ \alpha_0^2}+
n^2\left(\frac{\omega\sigma_{\xi}\sigma_{\alpha}}{\xi_0\alpha_0}\right)^2.
\end{equation}

\begin{figure}[hbt]
\centerline{\psfig{file=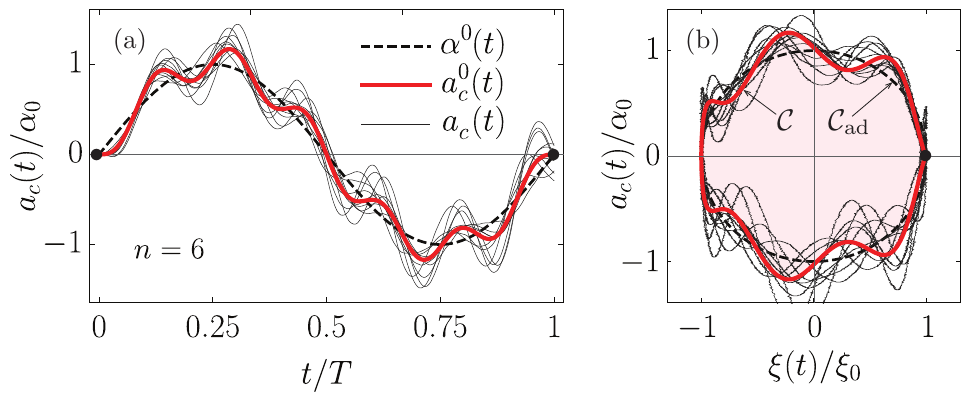,scale=.7}}

\vspace*{8pt}
\caption{Responses to circular driving with $n=6$ are shown. In (a) are  as functions of time shown noiseless driving $\alpha^0(t)$ (dashed line), noiseless response $a_c^0(t)$ (red) and $10$ responses $a_c(t)$ (black lines) to different realisations of white noise $\delta\alpha (t)$ with  intensity $\sigma_\alpha=\alpha_0/(20\sqrt{\omega})$. Bullets denote noiseless starting $[a_c^0(0),\xi^0(0)]$ and ending positions $[a_c^0(T),\xi^0(T)]$. In (b) are shown the same quantities as in (a) but as a function of coordinate driving $\xi(t)$ with $\sigma_\xi=\xi_0/(20\sqrt{\omega})$. The noiseless contours $\alpha^0[\xi^0]$ and $a_c^0[\xi^0]$ form closed loops, $\mathcal{C}_\mathrm{ad}$ and $\mathcal{C}$, respectively. Note that  $\phi$ is proportional to pink shaded area enclosed by  $\mathcal{C}$. In all calculations  $\alpha_0=\xi_0=1$ (in dimensionless units) was used.
}
\label{fig1}  
\end{figure}

In figure \ref{fig1}(a) are shown spin orbit responses as a function of time and in figure \ref{fig1}(b) is shown the contour ${\cal C}$ for the case of circular driving equation~\eqref{krog} with $n=6$. In both panels the dashed black lines denote noiseless spin orbit driving $\alpha^0(t)$ and the red line noiseless spin orbit response $a_c^0(t)$. The focus is on the set of $10$ spin orbit responses $a_c(t)$ to $10$ different  realisations of white noise in $\alpha(t)$. $\sigma_a^2(t)$ manifests as a spread of these curves around the ideal noiseless red line. Bullets correspond to initial $[a_c^0(0),\xi^0(0)]$ and final noiseless values  $[a_c^0(T),\xi^0(T)]$ of noiseless response and show that final values of $a_c(T)$ deviate from the desired ones. The noisy response is not periodic, resulting in open loop in parameter space unlike the case of noiseless $\mathcal{C}$ and noiseless adiabatic driving $\mathcal{C}_\mathrm{ad}$. Consequently the angle of spin rotation cannot be expressed as an area enclosed by the contour as in equation~\eqref{fi} and  in figure \ref{fig1}(b) pink shaded. It should be noted that in general the total angle of spin rotation $\phi$ is less prone to noise because the noisy curves oscillate around the ideal value and so contributions to final error partially cancel out.\cite{ulcakar17}

This  analysis of spin-rotation angle  demonstrated that  due to gate noise in the driving functions, spin transformations are not completely faithful. 
For non-adiabatic qubit manipulations the electron state is determined by the time-dependent Hamiltonian during the evolution and is in general a superposition of excited states, ultimately 
becoming the ground state when the transformation is complete. Therefore in addition to correct transformation of the spin direction, one has also to take care that the electron state has not left the starting energy manifold at the final time.  As shown in Refs.\cite{Cadez2013,Cadez2014,kregar16} such motions in parametric space can easily be performed if the driving functions are appropriately chosen. 
Here an important  question is relevant: how well does the final state of the electron relax to the desired final state energy manifold after the transformation if the driving function is not ideal as in the presence of noise?

In order to demonstrate how to answer this question in general we consider the qubit wave function  $|\Psi_{0\frac{1}{2}}(t)\rangle$, equation~\eqref{psi}, which is at $t=0$ in the ground state of the harmonic quantum dot (with $m=0$) and spin $\frac{1}{2}$. We observe its  relaxation to the ground state manifold that is spanned by two  basis states\cite{Cadez2013} of time dependent Hamiltonian equation~\eqref{H}  at time $t$,
\begin{equation}
|\widetilde\Psi_{0s}\rangle=e^{-im^{*}[x-\xi(t)]\alpha(t)\mathbf{n}\boldsymbol{\cdot\sigma}}|\psi_{0}[x-\xi(t)]\rangle|\chi_{s}\rangle.
\end{equation}
As the appropriate measure of the relaxation accuracy we define fidelity $F=\langle\Psi_{0\frac{1}{2}}(t)|P_0|\Psi_{0\frac{1}{2}}(t)\rangle$, where  $P_0=\sum_{s}|\widetilde\Psi_{0s}\rangle\langle\widetilde\Psi_{0s}|$ is the projector onto the ground state manifold.  We choose $\mathbf{n}$ perpendicular to the $z$-axis and a  straightforward derivation leads to 
the expression for overlaps of $|\Psi_{0\frac{1}{2}}(t)\rangle$ with the basis states at time $t$,
\begin{equation}
\langle\widetilde\Psi_{0\pm \frac{1}{2}}(t)|\Psi_{0\frac{1}{2}}(t)\rangle=\frac{1}{2}[e^{-\frac{1}{2}E_{+}(t)}\pm e^{-\frac{1}{2}E_{-}(t)}],
\end{equation}
where 
\begin{equation}
E_{\pm}(t)=\frac{m^{*}}{2\omega}\{[\omega(x_{c}(t)-\xi(t))\pm\dot{a}_{c}(t)/\omega]^{2}+[\dot{x}_{c}(t)\mp(a_{c}(t)-\alpha(t))]^{2}\}
\end{equation}
resembles classical energy with additional terms for spin-orbit coupling and is equal to the classical energy if the spin-orbit driving is constant.\cite{Cadez2013} Ideal qubit transformations with spin-fidelities ${\cal F}_s=|\langle\widetilde\Psi_{0s}|\Psi_{0\frac{1}{2}}\rangle|^2=\delta_{s\frac{1}{2}}$
are achieved  by applying ideal drivings, where the energies $E_\pm$ vanish at final time $t=T$, {\it i.e.}, when $x_c=\xi$, $a_c=\alpha$,  $\dot{x}_{c}=0$, and $\dot{a}_{c}=0$. 

The fidelity at arbitrary time $t$ is obtained by summation over final spin states,
\begin{equation}
F(t)=\sum_s {\cal F}_s(t)=\frac{1}{2}[e^{-E_{+}(t)}+e^{-E_{-}(t)}].
\end{equation}
The presence of noise in spin-orbit and spatial driving terms makes fidelity a random quantity, $F(t)=F^0(t)+\delta F(t)$, where $F^0(t)$ represents the result of noiseless driving and $\delta F(t)$ is the deviation from this value.  Fidelity is therefore
characterized by some probability density function $\frac{\mathrm{d}P(F)}{\mathrm{d}F}$. It 
can be calculated from the probability density for variables $E_{\pm}$ which  
are functions of independent random variables and normally distributed.
The probability density functions  for  $E_{\pm}$  can at time $t$ be calculated using the formula 
\begin{eqnarray}
\frac{\mathrm{d}P_\pm(E)}{\mathrm{d}E}{\Bigr\rvert_{t }}&=&\int\!\!\int\!\!\int\!\!\int
\delta[E-E_\pm(x_{c},\dot{x}_{c},a_{c},\dot{a}_{c})]
\\
& &\times
\frac{\mathrm{d}P_x(x_{c})}{\mathrm{d}x_{c}}\frac{\mathrm{d}P_{\dot{x}}(\dot{x}_{c})}{\mathrm{d}\dot{x}_{c}}
\frac{\mathrm{d}P_a(a_{c})}{\mathrm{d}a_{c}}\frac{\mathrm{d}P_{\dot{a}}(\dot{a}_{c})}{\mathrm{d}\dot{a}_{c}}
\mathrm{d}x_{c}\mathrm{d}\dot{x}_{c}\mathrm{d}a_{c}\mathrm{d}\dot{a}_{c}.\nonumber
\end{eqnarray}
The result is obtained by first calculating the characteristic functions, 
\begin{eqnarray}
p_\pm(k)&=&\int_{-\infty}^\infty \frac{\mathrm{d}P_\pm(E)}{\mathrm{d}E} e^{i k E} {\rm d}E={2\sigma_{1}^{-1}\sigma_{2}^{-1}\over\sqrt{(2  \sigma_1^{-2}-ik)(2\sigma_2^{-2}-ik)}},
\end{eqnarray}
with
\begin{eqnarray}
\sigma_{1}^{2}(t)&=&\left(\frac{2m^{*}}{\omega}\right)
[\omega^{2}\sigma_{x}^{2}(t)+\sigma_{\dot{a}}^{2}(t)/\omega^{2}],\\
\sigma_{2}^{2}(t)&=&\left(\frac{2m^{*}}{\omega}\right)
[\sigma_{\dot{x}}^{2}(t)+\sigma_{a}^{2}(t)].
\end{eqnarray}
Note the equality $p_+(k)=p_-(k)$ which after the inverse Fourier transform  yields equal functional forms for $E_+$ and $E_-$,
\begin{equation}
\frac{\mathrm{d}P_\pm(E_\pm)}{\mathrm{d}E_\pm}{\Bigr\rvert_{t }}=2\sigma_{1}^{-1}\sigma_{2}^{-1}I_{0}[(\sigma_{1}^{-2}-\sigma_{2}^{-2})E_\pm]e^{-(\sigma_{1}^{-2}+\sigma_{2}^{-2})E_\pm},
\end{equation}
where $I_{0}(z)$ is the modified Bessel function of the first kind.

\begin{figure}[ht]
\centerline{\psfig{file=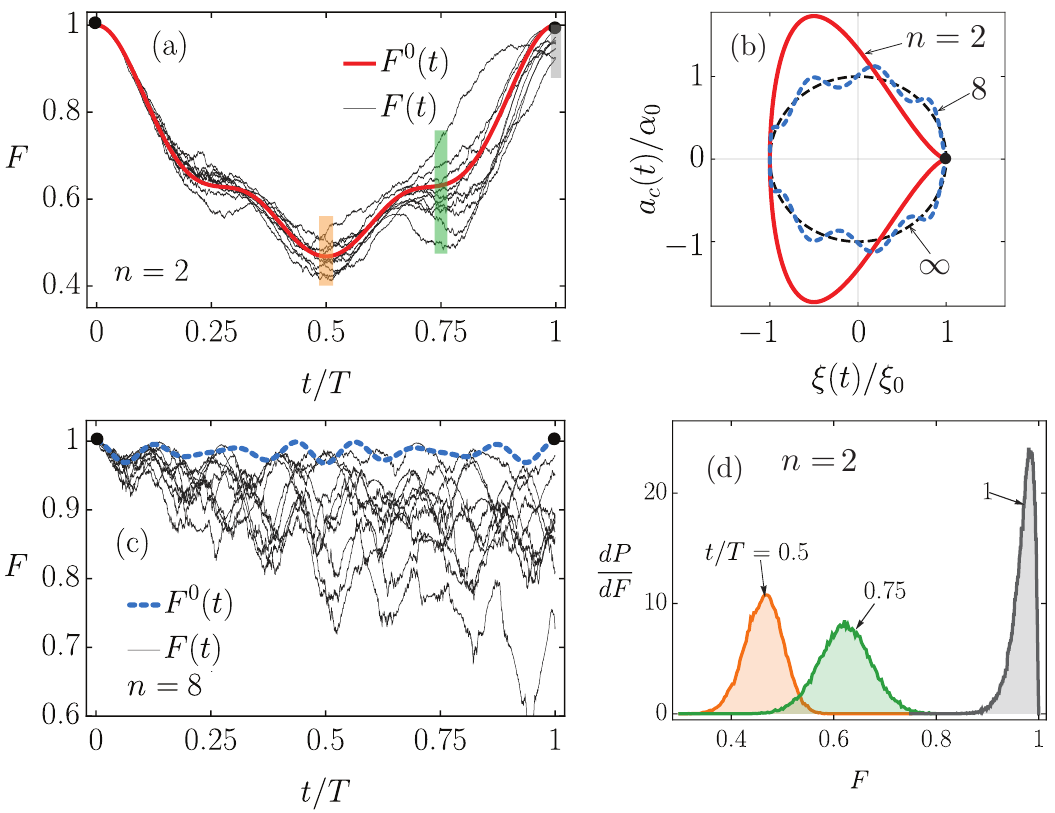,scale=0.7}}
\vspace*{8pt}
\caption{In panel (a) the noiseless fidelity $F^0(t)$ (red line) and $10$ fidelities $F(t)$ (black lines) for different realizations of white noise with intensities $\sigma_\xi/\xi_0=\sigma_\alpha/\alpha_0=1/(20\sqrt{\omega})$ are shown as functions of time when driving the system circularly with $n=2$ and $\alpha_0=\xi_0=1$ (in dimensionless units). Orange, grey and green shaded regions at times $t/T=0.5,\,0.75,\,1$, correspondingly, show the spread of noisy fidelities around the exact value. Noiseless contours in parameter space $[a_c,\xi]$ are for $n=2$ (red), $8$ (blue, dashed) and $n\to\infty$ (black, dashed) presented in panel (b). Panel (c) shows the same as panel (a) for circular driving with $n=8$ and the noiseless fidelity is denoted with dashed blue line. Bullets mark initial and final values of noiseless fidelity in (a) and (c) and contour in parameter space in (b). In panel (d) probability density distributions of fidelity for $n=2$ at times $t/T=0.5$ (orange), $t/T=0.75$ (green) and $t=T$ (grey) are shown. Colour codes coincide with the area of fidelity spreading shown in (a). Distributions were generated from $N=10^7$ samples.}
\label{fig2}
\end{figure}

Since the fidelity is a sum of two dependent random variables, its 
probability distribution is calculated from the joint probability distribution 
function for those two variables, which in general cannot be evaluated analytically. However, one can examine $\frac{\mathrm{d}P}{\mathrm{d}F}$ exactly
when $\sigma_{x}^{2}(t)=\sigma_{\dot{a}}^{2}(t)/\omega^{4}$ and $\sigma_{\dot{x}}^{2}(t)=\sigma_{a}^{2}(t)$, which is satisfied for $t=T$ if  the coordinate and the SOI driving noise intensities are equal, {\it i.e.}, 
$\sigma_\alpha=\omega\sigma_{\xi}$. 
In this case $E_{+}$ and $E_{-}$ become
{\it independent} random variables and $\frac{\mathrm{d}P}{\mathrm{d}F}$
can be calculated as the convolution of probability distributions for
$e^{-E_{+}}$ and $e^{-E_{-}}$. At $t=T$ the exact result  for $F\geq\frac{1}{2}$  is given by
\begin{equation}\label{eq:fidelity2d}
\frac{\mathrm{d}P(F)}{\mathrm{d}F}{\Bigr\rvert_{t =T}}=2\sigma_{F}^{-4}[B(\frac{1}{2F},\sigma_{F}^{-2},\sigma_{F}^{-2})-B(1-\frac{1}{2F},\sigma_{F}^{-2},\sigma_{F}^{-2})](2F)^{2\sigma_{F}^{-2}-1},
\end{equation}
where $B(x,a,b)$ is the incomplete beta function and $\sigma_{F}^{-2}=\sigma_{1}^{-2}+\sigma_{2}^{-2}$. 
For $F <\frac{1}{2}$ the probability distribution is $\frac{\mathrm{d}P}{\mathrm{d}F}=2\sigma_{F}^{-4}B(\sigma_{F}^{-2},\sigma_{F}^{-2})(2F)^{2\sigma_{F}^{-2}-1}$, where $B(a,b)$ is the beta function.
In practice where noise intensities are small the most relevant regime is $\sigma_{F}\to0$ for which the probability distribution equation~(\ref{eq:fidelity2d})  simplifies to  $\frac{\mathrm{d}P}{\mathrm{d}F} \propto (1-F)F^{2\sigma_{F}^{-2}}$. Due to similar dependence of ${\cal F}_s$ and $F$ on $E_\pm$ it is easy to derive analytical results also for spin-fidelity probability distributions $\frac{\mathrm{d}P_s}{\mathrm{d}{\cal F}_s}$ (not shown here).

In figure \ref{fig2}(a) different realizations of noisy fidelity (black lines) are compared to the noiseless one (red) for $n=2$. One can observe that noisy fidelity starts to deviate from noiseless one for $t/T\gtrsim0.1$, reaches maximum deviation at $t/T\sim0.5$ and then deviations are again lowered when approaching $t\to T$. The same quantities are  presented in figure \ref{fig2}(c) for circular driving with $n=8$ where noiseless curve is denoted with blue colour. Figure \ref{fig2}(b) shows noiseless curves $a_c[\xi^0]$ in parametric space  during the transformation with $n=2$ (red), $n=8$ (blue, dashed) and $n\to\infty$ (black, dashed), the latter corresponding to the adiabatic limit. Bullets denote initial and final values of $a_c(t)$ and $\xi(t)$. Note that the motion is  periodic with period $T$ and that  $a_c(0)=a_c(T)=\alpha(0)=\alpha(T)$ as is manifested  also in noiseless fidelity being equal to $1$ at $t=0$ and $t=T$, as a demonstration that the system returns  to the ground state manifold with probability $1$. This can be seen from positions of bullets in figures \ref{fig2}(a) and 2(c). Figure \ref{fig2}(d) shows the probability density distribution of fidelity at times $t/T=0.5$ (orange), $t/T=0.75$ (green) and $t=T$ (black). It should be mentioned that the distribution for $t=T$ is given also by exact formula, equation~(\ref{eq:fidelity2d}). The colour code of distributions corresponds to the code of the shading of fidelity spreading around the noiseless value in figure \ref{fig2}(a). Distributions are centred around noiseless values and their variances are proportional to spreadings observed in figure \ref{fig2}(a), the distribution at $t/T=0.5$ having the largest variance which is  lower at $t/T=0.75$ and even lower at $t=T$.

\section{Summary}

We presented an analysis of spin-qubit non-adiabatic manipulation of an electron traped in a moving linear harmonic trap and in the presence of time dependent Rashba interaction.  One of the main challenges here is a precise tuning of driving fields since the electron starting from the ground state should after performing one cycle with time-dependent Hamiltonian return to the ground state, although during the cycle the state of the electron is a superposition of excited eigenstates of the moving trap.

The problem is even more subtle because there will always be present some noise in driving functions, which means that spin-qubit transformation will always deviate from the ideal one. 
Since for the model considered here exact solutions are available for a broad class of drivings, we  concentrated also to the exact analysis of the influence of small deviations from ideal qubit manipulation. In particular,  we focused to an explicit example and
demonstrated how  one can analyse the effects of a general  noise to the transformation angle and we showed the results for the Ornstein-Uhlenbeck type of noise.

An example, considered in detail, is the case of circular driving in the space of parameters for which exact analytical formulae are given and analysed for white noise. In view of the fact that for non-adiabatic regimes a non-trivial point  is the ability of the system to return to the ground state after an arbitrary time-dependent driving, our analysis was focused to the  fidelity -- the overlap of the actual wave function with the desired ideal. For white noise  explicit formulae  are derived for symmetric noise intensities in position and spin-orbit driving functions. A detailed derivation and analysis of fidelity is presented. Additionally, analytical results are illustrated by special cases of driving together with numerically generated noisy drivings and the corresponding responses.

\section*{Acknowledgements}

The authors  acknowledge support from the Slovenian Research Agency under contract no. P1-0044.

\end{document}